\documentclass{pnastwo}

\usepackage{graphicx}
\usepackage{pnastwoF}
\usepackage{amssymb,amsfonts,amsmath}
\usepackage{multirow}
\usepackage{longtable}
\usepackage{appendix}
\usepackage{color}

\setlongtables

\url{}
\def \journalname{}
\footlineauthor{Keys, et al.}
\volume{}
\issuenumber{}

\begin{document}

\title{Calorimetric glass transition explained by hierarchical dynamic facilitation}

\author{
Aaron S. Keys\affil{1}{Chemical Sciences Division, Lawrence Berkeley National Laboratory, Berkeley CA, 94720}\affil{2}{Department of Chemistry, University of California, Berkeley CA, 94720},
Juan P. Garrahan\affil{3}{School of Physics and Astronomy, University of Nottingham, Nottingham, NG7 2RD, United Kingdom},\and
David Chandler\affil{1}{}\affil{2}{}$^1$}

\contributor{}

\maketitle

\begin{article}

\begin{abstract}
The glass transition refers to the \textcolor{black}{non-equilibrium} process by which an equilibrium liquid is transformed to a non-equilibrium disordered solid, or vice versa. \textcolor{black}{Associated response functions, such as heat capacities, are markedly different on cooling than on heating}, and the response to melting a glass depends markedly on the cooling protocol by which the glass was formed. This paper shows how this irreversible behavior can be interpreted quantitatively in terms of an East-model picture of localized excitations (or soft spots) in which molecules can move with a specific direction, and from which excitations with the same directionality of motion can appear or disappear in adjacent regions.
As a result of this facilitated dynamics, excitations become correlated in a hierarchical fashion.  These correlations are manifested in the dynamic heterogeneity of the supercooled liquid phase.  While equilibrium thermodynamics is virtually featureless, a non-equilibrium glass phase emerges when the model is driven out of equilibrium with a finite cooling rate.  The correlation length of this emergent phase is large and increases with decreasing cooling rate.  A spatially and temporally resolved fictive temperature encodes memory of its preparation.  Parameters characterizing the model can be determined from reversible transport data, and with these parameters, predictions of the model agree well with irreversible differential scanning calorimetry.

\end{abstract}

\keywords{dynamic heterogeneity | supercooled liquids | heat capacity | differential scanning calorimetry}
\abbreviations{DSC, differential scanning calorimetry; TNM, Tool-Narayanaswami-Moynihan; KCM, kinetically constrained model}

\dropcap{W}hen a supercooled liquid is cooled significantly below its glass transition temperature, the system freezes into an amorphous vitrified state~\cite{ediger1996supercooled, angell1995formation}.  This process encodes a unique non-equilibrium structural signature within the glassy material, a signature that is not apparent in equilibrium correlation functions, but is evident from thermodynamic properties and response functions, such as the enthalpy, specific volume, heat capacity or thermal expansivity.  The complex temperature variation of these properties for systems heated from the vitrified state reveals that glassy materials bear a distinct memory of the protocol by which they were created.  Here, we show how this irreversible behavior can be understood in terms of the non-equilibrium behavior of a model we have used in the past to treat reversible behaviors of glass-forming liquids~\cite{eastmodel, garrahan2002geometrical, garrahan2003coarse, chandler2010dynamics}. In so doing, we uncover an underlying dynamical transition between equilibrium melts with no trivial spatial correlations and non-equilibrium glasses with correlation lengths that are both large and dependent upon the rate at which the glass is prepared. 

\begin{figure} [b]
\centerline{\includegraphics[width=0.49\textwidth]{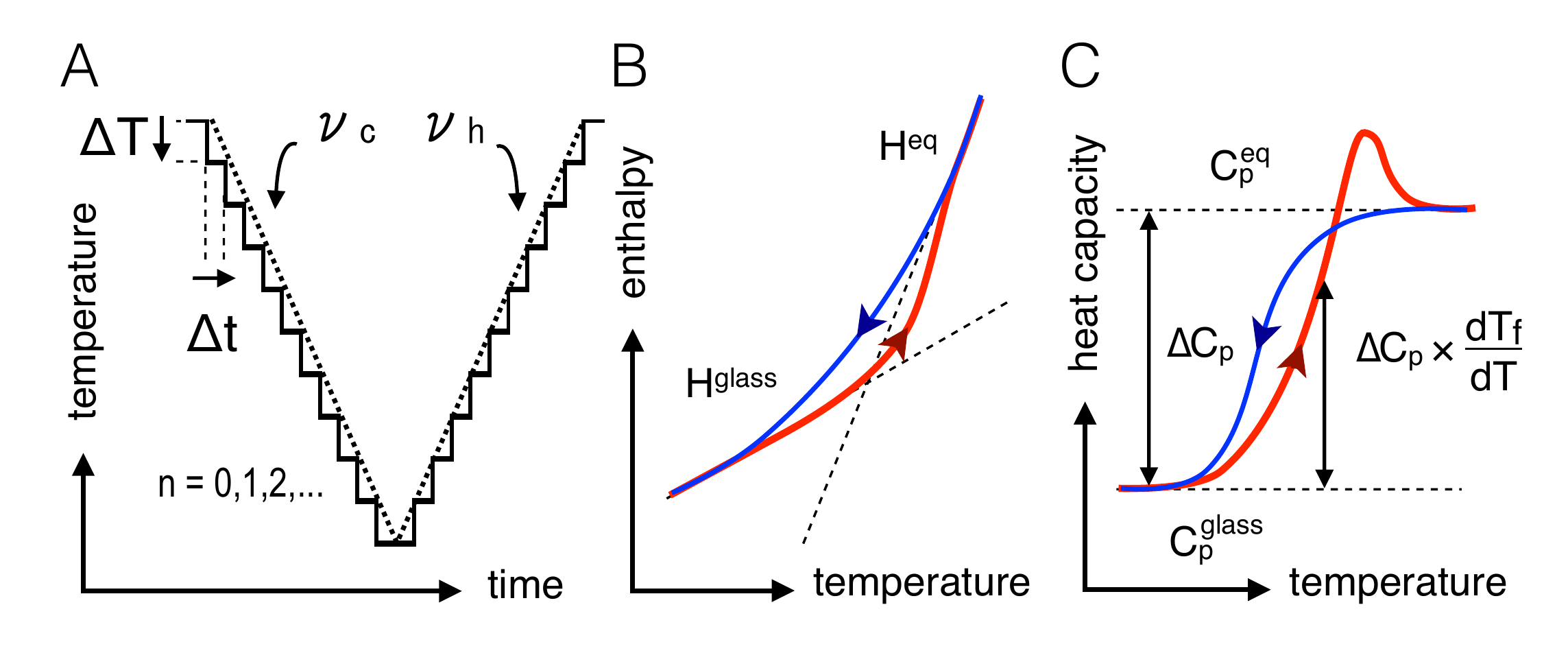}}
\caption{\label{fig:fig1}  {Schematics of a standard temperature cycle ({\it A}) and corresponding thermodynamic quantities ({\it B}) and ({\it C}). \textcolor{black}{For theoretical analysis, a continuously changing temperature $T$ is treated as a discrete sequence}, $k=1,2,\cdots$, each step having time duration $\Delta t$.  The working temperature at the $(k+1)$th step differs from that at the $k$th step by an amount $\Delta T$.  The rates of cooling and heating $\nu \equiv \Delta T / \Delta t$ are denoted $ \nu_\mathrm{c}$ and $ \nu_\mathrm{h}$, respectively. ({\it B}) Enthalpy $H$, and ({\it C}) the corresponding heat capacity $C_\mathrm{p}$, for a system that is cooled and then heated through the glass transition (blue, red).  The irreversible glass transition occurs over the range of temperatures where the blue and red curves differ.  A standard definition of fictive temperature, $T_\mathrm{f}$, is given in terms of its temperature derivative in that range: $C_\mathrm{p} = C_\mathrm{p}^{\rm glass} +\Delta C_\mathrm{p} (dT_\mathrm{f}/dT) $, where $\Delta C_\mathrm{p} \equiv C_\mathrm{p}^{\rm eq}-C_\mathrm{p}^{\rm glass}$.  As the system tends to equilibrium, $T_\mathrm{f} \rightarrow T$.
}}
\end{figure}

The class of experimental protocols we consider is illustrated in Fig.~\ref{fig:fig1}{\it A}.  The temperature $T$ is a function of time with a rate of temperature change, $\nu = \Delta T / \Delta t$, that is negative for cooling ($\nu = \nu_\mathrm{c} <0$) and positive for heating ($\nu = \nu_\mathrm{h} > 0$) .  The structural relaxation time of the liquid, $\tau$, increases as temperature decreases.  At a sufficiently low temperature, $|d\tau/dT|$ exceeds $|1/\nu_\mathrm{c}|$.  At this stage, the liquid begins to fall out of equilibrium, forming a glass.  The process is continuous but precipitous.  A glass transition temperature, $T_\mathrm{g}$, is defined in various ways, but always such that it is a temperature in this precipitous region. Heat capacities measured by differential scanning calorimetry (DSC)~\cite{moynihan1976structural, hodge1994enthalpy} and corresponding enthalpies are typical experimental indicators of the transition.  Their behaviors are illustrated schematically in Figs.~\ref{fig:fig1}{\it B},{\it C}.  

A most important feature illustrated in the figure is time asymmetry.  Cooling produces a monotonic decrease in heat capacity, while heating produces a non-monotonic and anomalous response, the size of which (we discuss below) depends upon the rate at which the glass was produced by cooling.  Hysteresis between equilibrium phases lacks this asymmetry.   We do not focus on the end points of those curves -- the heat capacities of the glass and liquid phases, $C_\mathrm{p}^\mathrm{glass}$ and $C_\mathrm{p}^\mathrm{eq}$, respectively.  The difference, $\Delta C_\mathrm{p} =C_\mathrm{p}^\mathrm{eq}-C_\mathrm{p}^\mathrm{glass}$, one may argue~\cite{trachenko2011heat}, is decoupled from dynamics and understood in terms of time-independent elastic responses of the two phases~\cite{trachenko2011heat, bolmatov2012phonon}.  In contrast, the irreversible transitional behaviors found upon cooling and heating can only be understood in terms of dynamics.  It is these irreversible behaviors we consider here. 

To do so, we generalize the concept of fictive temperature~\cite{tool1931variations, tool1946relation} to a spatially resolved field, $T_\mathrm{f}(\vec{r}, t)$.  To define this quantity and describe its utility, note that a time-dependent working temperature $T(t)$ (i.e., the temperature of a surrounding bath) equilibrates quickly throughout a glass or glass-forming material  -- on a time scale that is effectively instantaneous in comparison to structural equilibration times.  Further, structural relaxation occurs locally and intermittently.  A short time after a local region surrounding position $\vec{r}$ happens to reorganize, say at time $t'$, the configurational contribution to a local thermodynamic property, like an enthalpy density $h(\vec{r},t)$, will be locked at \textcolor{black}{a value, $h_\mathrm{eq}(T(t'))$, typical of the equilibrium distribution of values for temperature $T(t')$, and it will change} only after the region reorganizes again.  Therefore, we define $T_\mathrm{f}(\vec{r}, t)$ to be the working temperature of the system during the most recent structural reorganization of the region surrounding $\vec{r}$.  As such, 
extensive thermodynamic properties, like configurational enthalpy $H$, can be written as 
\begin{equation}
H(t) = \int \mathrm{d} {\vec{r}}\,\, h_\mathrm{eq}(T_\mathrm{f}(\vec{r}, t))\,.
\label{eq:enthalpy}
\end{equation}
The net enthalpy is this configurational part plus a regular contribution from molecular vibrations.  The latter is simply characterized by the working temperature. 

Tool's original formulation~\cite{tool1931variations, tool1946relation} is similar, but without considering spatial variation.  In reality, different domains within the system lose and regain equilibrium at different temperatures.  Thus, the non-equilibrium distribution of particle arrangements is never representative of an equilibrium configuration at any temperature~\cite{tool1946relation, ritland1956limitations, moynihan1976dependence}.  Phenomenological improvements to Tool's original approach~\cite{hodge1994enthalpy}, including the Tool-Narayanaswami-Moynihan (TNM) model~\cite{narayanaswamy1971model, moynihan1976structural}, introduce equations of motion for the fictive temperature, but still in mean field. These models require extensive parameterization to correct for neglected microscopic details.  \textcolor{black}{Improvements, e.g., Refs~\cite{kovacs1979isobaric, lubchenko2004theory, richert2011heat, mazinani2012enthalpy}, offer phenomenological approaches that go} beyond mean field, but still with many parameters and without a concrete microscopic model for dynamics.

In contrast, we consider a spatially-resolved fictive temperature field that is coupled to an underlying dynamics exhibiting fluctuations on all but the smallest trivial scale.  Specifically, we employ ``East-like'' models.  The original East model~\cite{eastmodel} is one of many kinetically constrained models (KCMs) -- models that have been invented to study glassy dynamics in idealized contexts\cite{fredrickson1984kinetic, ritort2003glassy}.  The East model, in particular, treats a one-dimensional lattice, with excitations distributed at random on the lattice stretching from from right (``east'') to left (``west'').  The dynamics of the model is non-trivial because the birth and death of an excitation requires the presence of another excitation immediately to its east.  At low enough excitation concentrations, this directional facilitation leads to hierarchical dynamics~\cite{palmer1984models}, where relaxation over a domain of length $\ell$ requires an energy $J_\ell$ that grows logarithmically with $\ell$~\cite{aldous2002asymmetric}.  It thus encodes dynamical behavior posited by Palmer, et al.~\cite{palmer1984models}.  

Generalizations of the East model to higher dimensions exhibit similar scaling.  Most important, the behaviors of these East-like models -- directional facilitation and concomitant logarithmic scaling -- have been shown to emerge from molecular dynamics of atomistic models of glass forming liquids~\cite{keys2011excitations}, and also, this scaling has been shown to collapse the seemingly disparate results of thousands of measurements of reversible structural relaxation in supercooled melts~\cite{elmatad2009corresponding}.  The correspondence between behaviors of real materials and East-like behavior follows from the two essential features of glass-forming liquids.  First, the possibility of local reorganization is coincident with a local soft spot, and small movements within such a region can lead to a softening of an adjacent region~\cite{keys2011excitations}; this feature is the origin of facilitation. Second, the direction of a displacement or movement is preserved by the movements in the newly softened region~\cite{keys2011excitations}; this feature is the origin of directionality or string-like motions~\cite{donati1998stringlike}.  East-like models contain both features, and essentially nothing else.

\subsection{East model illustrates fictive temperature, domain size and glassy memory}

To illustrate how East-like models work, we consider first the original East model with $N$ lattice sites, $i=1,2,\cdots, N$, each of which can be in one of two states, $n_{i} = 0$ or $1$.  The latter represents an excited site.  Site $i$ can excite or de-excite provided the adjacent neighbor to the ``east'' is excited. This dynamics is encoded in the transition rates for that site, $k_{i, 0 \rightarrow 1} =  n_{i-1} c/(1-c)$ and $k_{i, 1 \rightarrow 0} = n_{i-1}$, where $c \equiv \langle n_{i} \rangle$, and $c/(1-c) = \exp(-1/T)$.  We use units where excitation energy over Boltzmann's constant is unity, and the time of a single integration step is unity. The spatial variable, $\vec{r}$, is the dimensionless lattice-site label $i$.   

The transition rates obey detailed balance, so the model relaxes to its equilibrium state.  In this state, excitations are distributed at random, where $P(\ell) =c \,\exp(-c \, \ell)$ is the probability that a tagged excitation is separated from its nearest excited neighbor by $\ell$.  Integration of the stochastic dynamics for this model can be done in various ways ~\cite{ashton2005fast}.  For the results shown in this section it is carried out with a continuous time algorithm~\cite{gillespie1977exact} starting from an equilibrated high-temperature state.  Each cooling or heating step in a cycle depicted in Fig.\ref{fig:fig1}{\it A} is propagated for many integration-time steps up to the time $\Delta t$.  The next cooling or heating step proceeds from the configuration obtained at the end of the previous step, but with the temperature shifted by the amount $\nu \Delta t$. 

While the equilibrium behavior is unstructured, the dynamics of the model becomes notably heterogeneous below the onset temperature, $T_\mathrm{o} \approx 1$.   In that low temperature regime, the structural relaxation time is non-Arrhenius, obeying the parabolic law~\cite{elmatad2009corresponding}, 
\begin{equation}
\ln(\tau / \tau_\mathrm{mf}) = J^2(1/T - 1/T_\mathrm{o})^2\,, \quad T < T_\mathrm{o},
\label{eq:parabolic}
\end{equation}
where $J^2 =1/2\ln2$~\cite{aldous2002asymmetric,chleboun2012time}, and $\tau_\mathrm{mf}$ is the structural relaxation time in mean field, which has Arrhenius temperature dependence and is an accurate approximation for $\tau$ when $T \gtrsim T_\mathrm{o}$.  The universal super-Arrhenius behavior, Eq.~\eqref{eq:parabolic}, is found in real systems for $T<T_\mathrm{o}$, but with different $J$ and different $\tau_\mathrm{mf}$.  (In most glass forming liquids, the temperature variation of $\tau_\mathrm{mf}$ is negligible~\cite{elmatad2009corresponding}.)

The universal behaviors of fluctuations or dynamic 
heterogeneities for $T<T_\mathrm{o}$ can be quantified with a persistence field~\cite{jung2005dynamical}, $p_i(t,t') =  \delta \left[ \sum_{t''=t' }^{t} \kappa_i(t'') \right]$.  Here, $\delta[X]$ 
stands for Kronecker's delta, which is 1 when $X=0$, and zero otherwise, and $\kappa_i(t)= | n_i(t) - n_i(t-1) |$ indicates whether a change in state (or kink) has occurred at site $i$ and time $t$.  Thus, the persistence field at site $i$, $p_i(t,t')$, is 1 if and only if there has been no dynamical activity at site $i$ over the time frame between $t'$ and $t$. This quantity provides a recursive relationship for the fictive temperature field, $T_{\mathrm{f},i}(t)$.  Specifically, with working temperature $T(t)$,
\begin{equation}
\label{eq:Tfi}
T_{\mathrm{f},i}(t) = p_i(t,t - \Delta t) T_{\mathrm{f},i}(t - \Delta t) + \left[ 1-p_i(t,t - \Delta t) \right] T(t) .
\end{equation}
When the East model is driven out of equilibrium with a temperature that changes in steps of time duration $\Delta t$, the three fields together,  $n_i(t)$, $p_i(t,t - \Delta t)$  and $T_{\mathrm{f},i}(t)$, elucidate the resulting dynamics.

\begin{figure} [b]
\centerline{\includegraphics[width=0.5\textwidth]{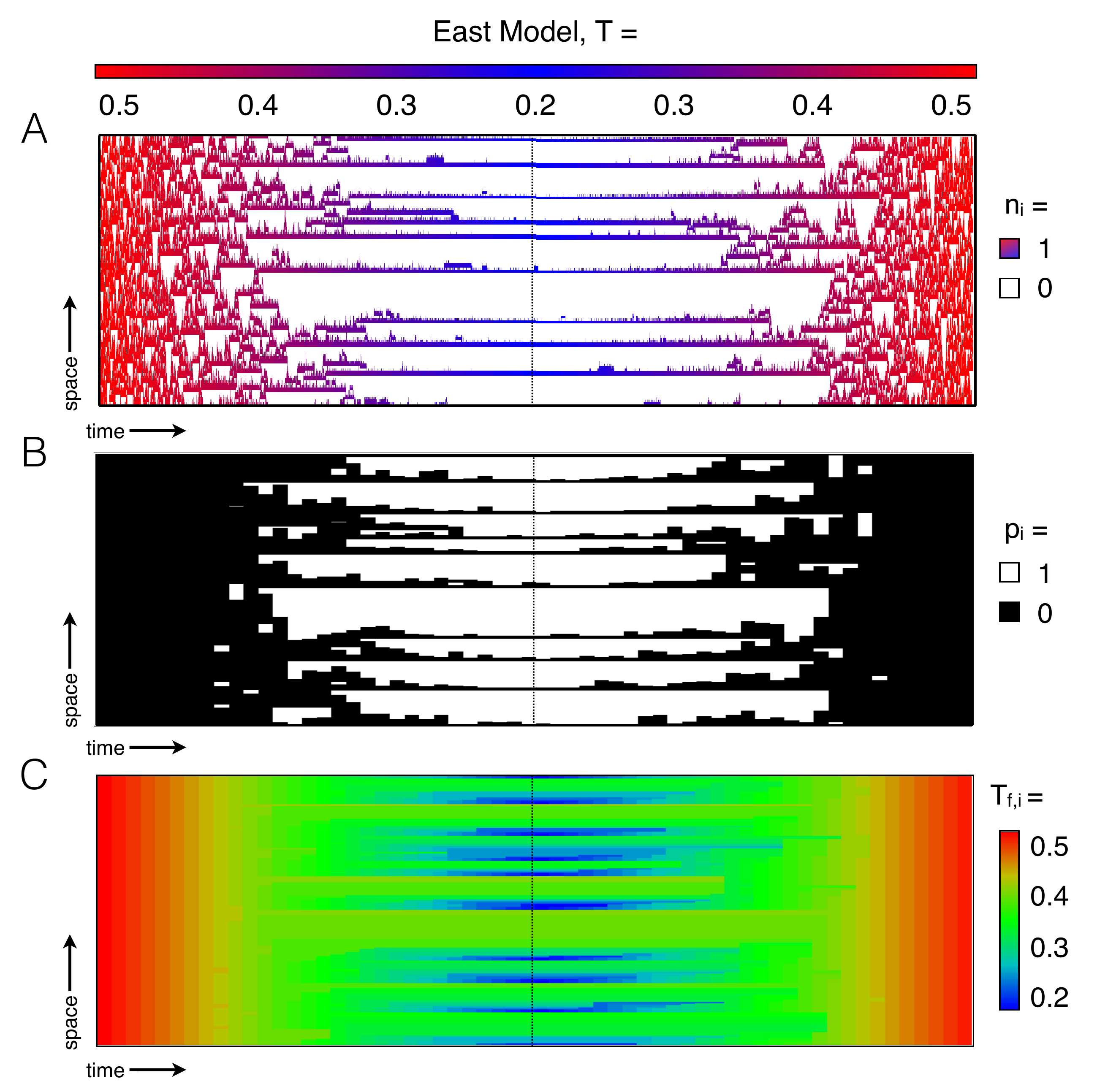}}
\caption{\label{fig:fig2}  {Cooling and heating cycle for the East model and the emergence of a non-equilibrium striped phase.  For this illustrated trajectory, the cooling and heating rates are $ |\nu_\mathrm{c}| = |\nu_\mathrm{h}| = 10^{-7}$, which is the value of  $(\mathrm{d}\tau / \mathrm{d}T)^{-1}$ for the East model at $T\approx 0.37$.  Panel ({\it A}) depicts the evolution of excitations as a function of temperature.  Sites are colored according to the external temperature $T(t)$.  Upon heating, the system retains memory of the cooling protocol, manifested by a striped configuration containing similarly-sized domains.  Panel ({\it B}) depicts the corresponding persistence field.  Sites that persist during the $k$th step are colored white while sites that relax are colored black.  Panel ({\it C}) depicts the fictive temperature field derived from the persistence field.  The color scheme indicates the value the instantaneous fictive temperature at each lattice site.}}
\end{figure}

\begin{figure}
\centerline{\includegraphics[width=0.5\textwidth]{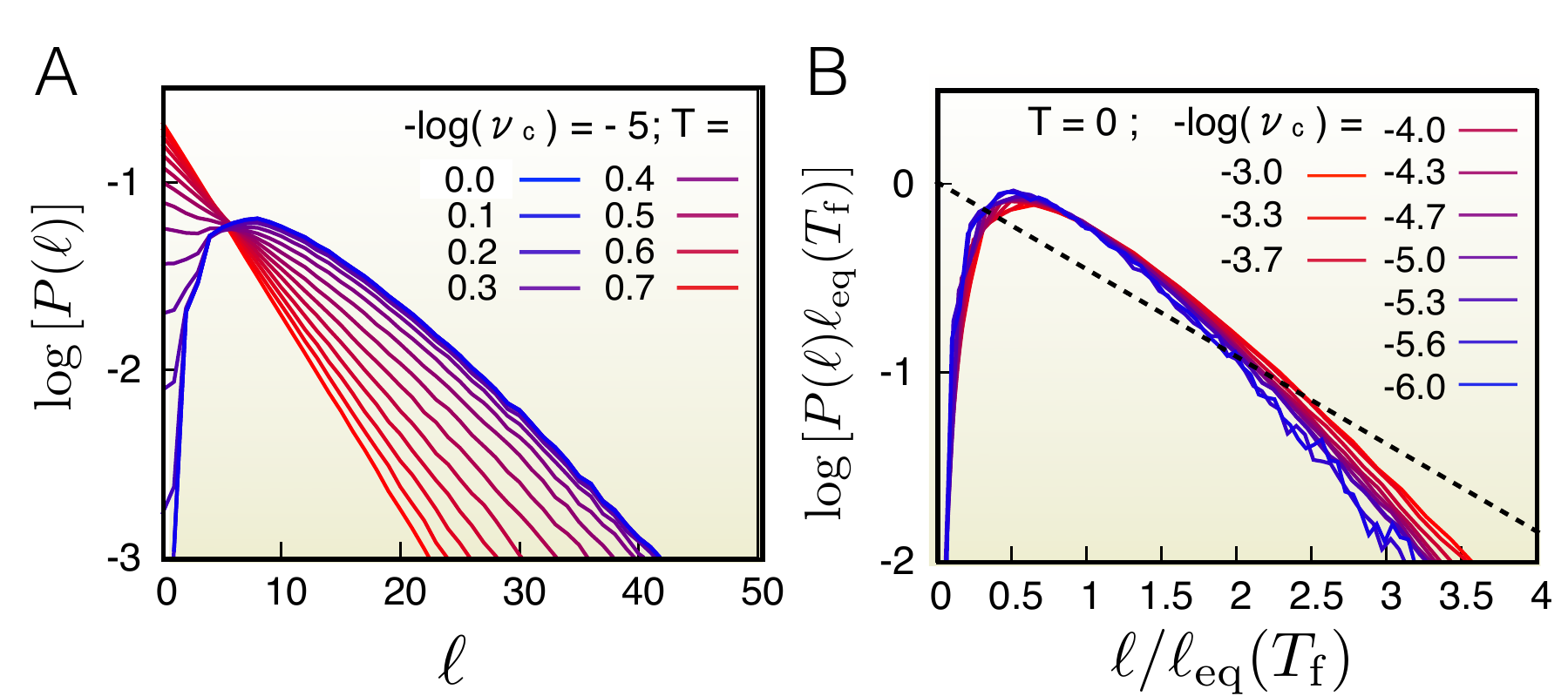}}
\caption{\label{fig:fig3}  {Size distributions of glassy domains in the East model after cooling with frequency $\nu_\mathrm{c}$ to the temperature $T$.  Statistics obtained by averaging over many trajectories like the one illustrated in Fig.\ref{fig:fig1}.  Panel ({\it{A}}) shows the probability distribution of domain sizes $P(\ell)$ for excitations in the East model for a cooling rate of $\nu_c = -10^{-5}$.   Above the kinetic glass transition temperature $T_\mathrm{g} \approx 0.48$, excitations obey ideal gas statistics with $P(\ell)\propto \exp(-c\ell)$.  Below $T_\mathrm{g}$, excitations develop pair correlations with the most probable domain size occurring near $\ell = 10$.  As the cooling rate slows, this characteristic domain size grows.  Panel ({\it{B}}) shows $P(\ell)$ for glasses formed by the East model at different cooling rates, all quenched to $T=0$.  The curves collapse when scaled by $\ell_\mathrm{eq}(T_\mathrm{f})$, the equilibrium length corresponding to the spatially averaged fictive temperature of the glass.  The black dashed line shows the variation with $\ell$ for the equilibrium distribution, $P_\mathrm{eq}(\ell) \propto \exp(-\ell/ \ell_\mathrm{eq})$.}}
\end{figure}

Panels {\it{A}}, {\it{B}} and~{\it{C}} of Fig.~\ref{fig:fig2} depict these fields for a system cooled and re-heated through $T_\mathrm{g}$.  At the beginning and end of the trajectories, excitations initially assume a gas-like distribution.  See Fig.~\ref{fig:fig3}{\it{A}}.  In that regime, trajectories show space-time bubbles characteristic of reversible dynamics of a glass former~\cite{garrahan2002geometrical, chandler2010dynamics}.  In the high-temperature regime, the system is at equilibrium, and the \textcolor{black}{ distribution of inter-excitation distances is exponential with $\ell = 0$ most probable. 
As temperature is lowered, excitations annihilate to adjust to the lower equilibrium concentration, and the average domain size grows, but still $\ell=0$ is the most probable.}  Because the activation energy to equilibrate a domain, $J_\ell$, increases with $\ell$, relaxation occurs more readily for excitations separated by smaller distances than for larger distances, a relationship we quantify in the following section.  

At low temperatures, when the system begins to fall out of equilibrium, it does so over a series of stages during which ever-smaller domains freeze into place.  Due to the hierarchical nature of the dynamics, therefore, smaller domains acquire lower fictive temperatures.  The result is seen in the complex spatial pattern for the fictive temperature field $T_{\mathrm{f},i}$ illustrated in Fig.~\ref{fig:fig2}{\it C}.  The frozen phase that eventually forms, corresponding to glass, is a striped phase in space-time.  It contains an excess of large domains relative to that of the equilibrium phase, corresponding to melt or liquid.   This result is quantified with the distributions shown in Fig.~\ref{fig:fig3}.  After the formation of the glass, i.e., below $T_\mathrm{g}$, the distribution peaks at a characteristic value of \textcolor{black}{ $\ell > 0$}.  This length is the typical size of the smallest domains that cannot relax during the cooling process.  It increases as cooling rate decreases.  Fig.~\ref{fig:fig3}{\it{B}} shows that the distributions collapse when scaled by the equilibrium domain length, $\ell_\mathrm{eq}(T) = \exp(1 / T)$, evaluated at the spatially averaged fictive temperature, $T_\mathrm{f} = (1/N)\sum_i T_{\mathrm{f}, i}$. 

Thus, there is a structural distinction between the glass and its melt.  The former has a non-trivial static length that results from an irreversible dynamical process.  The latter has no non-trivial static length. The frozen system and its equilibrium counterpart can have the same excitation concentration, $c$, but with different spatial distributions of excitations.    
An instantaneous quench \textcolor{black}{to $T = 0$} will not produce this distinct structural behavior of the non-equilibrium system.  \textcolor{black}{A finite cooling rate or physical aging~\cite{sollich1999glassy, sollich2003glassy} below $T_\mathrm{g}$ is required to produce this transition to a glass.}  An instantaneous quench of an equilibrium system will remove surging fluctuations surrounding anchored excitations, but will otherwise leave the spatial distribution of excitations identical to that of the equilibrium system.  The quenched system will thus melt as soon as temperature is returned to any finite value, equilibrating within a reversible relaxation time at that temperature.  \textcolor{black}{Precursors to the glass phase with finite correlation lengths emerge out of equilibrium after a quench to a low but \emph{finite} temperature~\cite{sollich1999glassy, sollich2003glassy}.}

Lowering cooling rate produces larger glassy domain sizes, and thus lowering cooling rate produces more stable glass.  It is another manifestation of hierarchical dynamics (i.e., $J_\ell$ growing with increasing $\ell$).  An excitation can spawn new excitations, building lines of excitations typically as long as $\ell_\mathrm{eq}(T)$.  Longer lines will be exponentially rare.  Thus, for $T < T_\mathrm{g}$, excitations are isolated and cannot relax because $\ell_\mathrm{eq}(T)\ll \ell_\mathrm{eq}(T_\mathrm{f})$.   As the system is heated from its frozen state, minor surges of activity can occur on short length scales, which causes the fictive temperature to increase slightly.  But the glassy domains persist until, at warm enough temperatures, the system finally acquires sufficient energy to relax the domains of characteristic size.  At this point equilibration occurs rapidly.  We will see shortly, that this behavior is the essence of the asymmetric responses characteristic of the calorimetric glass transition. 

Hysteresis \textcolor{black}{in response to a time-varying field} is a general property of any system with dynamical bottlenecks \textcolor{black}{or separations in time scales}.  Thus, it will be a property of all KCMs (e.g., Refs.~\cite{franz1995dynamical, brey2000thermodynamic}).  But the time-asymmetry or encoded memory of preparation discussed here is a property of hierarchical dynamics~\cite{prados2001glasslike} \textcolor{black}{as explained in the previous paragraphs}.  Domains in KCMs that have diffusive (i.e., non hierarchical) dynamics, such as the 1-spin facilitated Fredrickson-Andersen model~\cite{fredrickson1984kinetic} and the Backgammon model~\cite{ritort2003glassy}, tend to evolve toward equilibrium upon heating is much like that of standard unconstrained systems (e.g., the Ising model~\cite{brey1994dynamical}). Hierarchical KCMs such as the East model, in contrast, must populate lower energy levels before relaxing larger domains.  \textcolor{black}{Further discussion and comparison between prior approaches and models are given in \textit{Supporting Information} (\textit{SI}).}

\subsection{Comparison with DSC Experiments} Differential scanning calorimetry (DSC) determines a non-equilibrium heat capacity, or more precisely, the rate of change of enthalpy as a function of time.  It is linked to the fictive temperature through Eq.~\eqref{eq:enthalpy}.  To a good approximation, the equilibrium density of configurational enthalpy is a linear function of temperature over the relevant range of temperatures.    With that approximation, the non-equilibrium configurational heat capacity is $C_\mathrm{p} = C_\mathrm{p}^{\mathrm{glass}} + \Delta C_\mathrm{p} \,\nu^{-1} \,   dT_\mathrm{f}/dt$, i.e.,  
\begin{equation}
\tilde{C}_\mathrm{p}(T) \equiv [C_\mathrm{p}(T) - C_\mathrm{p}^\mathrm{glass}]/\Delta C_\mathrm{p} = \mathrm{d} T_\mathrm{f} /\mathrm{d} T,
\label{eq:tildeC}
\end{equation}
where $\Delta C_\mathrm{p}$ is the difference between liquid and glass heat capacities outside the glass-transition region, and $T$ refers to the working temperature at the time when the non-equilibrium heat capacity is measured.  The spatially averaged fictive temperature, $T_\mathrm{f}$, is a functional of $T(t)$ encoding preparation history.  Its working-temperature derivative in Eq.~\eqref{eq:tildeC} is evaluated at the measurement time.  [Expressions more complicated than \eqref{eq:tildeC} apply when $h_\mathrm{eq}(T)$ is a non-linear function of $T$ and when the working temperature is a non-linear function of $t$.]

The original East model requires generalization to account for dimensionality and differing energy scales of different experimental systems.   The logarithmic growth of activation energy for domains of length $\ell$, is written as~\cite{keys2011excitations}
\begin{equation}
\label{eq:Jell}
J_\ell = J_\mathrm{\sigma} \left[1 + \gamma \ln \left(\ell / \sigma\right) \right]\,,
\end{equation}
where $\sigma$ stands for a principal structural length in the system (e.g., a molecular diameter), and $J_\sigma$ is the activation energy for displacements or reorganization on that scale.  For the original East model $\sigma = 1$, $J_\sigma =1$ and $\gamma =1/2 \ln2$ sets the effective energy barrier~\cite{chleboun2012time}.  More generally, $\sigma$, $J_\sigma$ and $\gamma$ are material properties.  We use the terminology ``East-like'' to refer to this general class of models. From Eq.\eqref{eq:Jell}, the time scale to relax a domain of length $\ell$ is 
\begin{equation}
\label{eq:tauell}
\tau_\ell = \tau_\mathrm{o} \, \exp\left[ \left( 1/T- 1/T_\mathrm{o}\right)  \left(J_\ell - J_\mathrm{\sigma}\right) \right]\,, 
\end{equation}
where $\tau_\mathrm{o}$ is the mean-field relaxation time at the onset temperature.  (We neglect the temperature dependence in $\tau_\mathrm{mf}$, \textcolor{black}{and we consider only $T<T_\mathrm{o}$}.)  The parabolic law, Eq.\eqref{eq:parabolic}, is a consequence because the equilibrium domain size grows exponentially with $1/T$.  Specifically,
\begin{equation}
\label{eq:elleq}
\ell_\mathrm{eq} = \sigma \exp\left[ \left( 1/T- 1/T_\mathrm{o}\right) J_\mathrm{\sigma}/d_\mathrm{f} \right]\,,
\end{equation}
where $d_\mathrm{f}$ is the fractal dimensionality of the dynamic heterogeneity.  This gives $J = J_\sigma \sqrt{\gamma / d_\mathrm{f}}$.  The original East model in dimension $d$ has $d_\mathrm{f} = d$~\cite{ashton2005fast}.  

Thus, when employing an East-like model to interpret experimental results for $\tilde{C}_\mathrm{p}$ values for $J$, $T_{o}$, and $\tau_\mathrm{o}$ are required.  These are known from reversible transport data~\cite{elmatad2009corresponding}.   Values for ${\sigma}$, $\gamma$, and $d_\mathrm{f}$ are also required.  Numerical studies suggest that the last of these is universal, depending only upon the dimensionality and symmetry of the system~\cite{keys2011excitations}.  As such, we take its value as that computed from three-dimensional atomistic models glass formers:  $d_\mathrm{f} \approx 0.8\, d = 2.4$~\cite{keys2011excitations}.  These numerical studies also find system-dependent values of $\gamma$ ranging from 0.2 to 0.5.  From a scaling argument (see {\it Methods}), $\gamma$ and $\sigma$ can be collapsed to a reference value, $\gamma_0 \approx 0.275$, and a reduced length $\sigma / \ell_0$ that is of order 1.  Therefore, each different system is treated with one fitting parameter, either $\sigma/ \ell_0$ or, equivalently, $\gamma$.  The parameters employed are collected in Table~1.  
 
For the relevant temperature regime, East-like dynamics can be treated by quadrature~\cite{sollich1999glassy, sollich2003glassy}.  The procedure is an iterative algorithm derived from the fact that during a cooling or heating step of duration $\Delta t$, there is an upper limit to equilibrated domain size, $\Delta \ell$.  From Eqs.~\eqref{eq:Jell} and \eqref{eq:tauell}, substituting $\Delta t$ for $\tau_\ell$ and $\Delta \ell$ for $\ell$,
\begin{equation}
\label{eq:DeltaEll}
\Delta \ell /\sigma = \left(\Delta t / \tau_\mathrm{o}\right)^{1 / \left[ (1/ T - 1/T_\mathrm{o}) \gamma J_\mathrm{\sigma}\right]}.
\end{equation}
This relationship prescribes a method for computing the non-equilibrium persistence and fictive temperature fields for an experimental system by relaxing domains smaller than $\Delta \ell$ at each temperature step, employing the Sollich-Evans superdomain construction~\cite{sollich2003glassy} to properly account for domain regeneration during the relaxation process (see \textit{Methods}). The domain regeneration ensures that an equilibrium ensemble is maintained when the system is ergodic and $T$ is not varying with time.  

\begin{figure}[t]
\centerline{\includegraphics[width=0.5\textwidth]{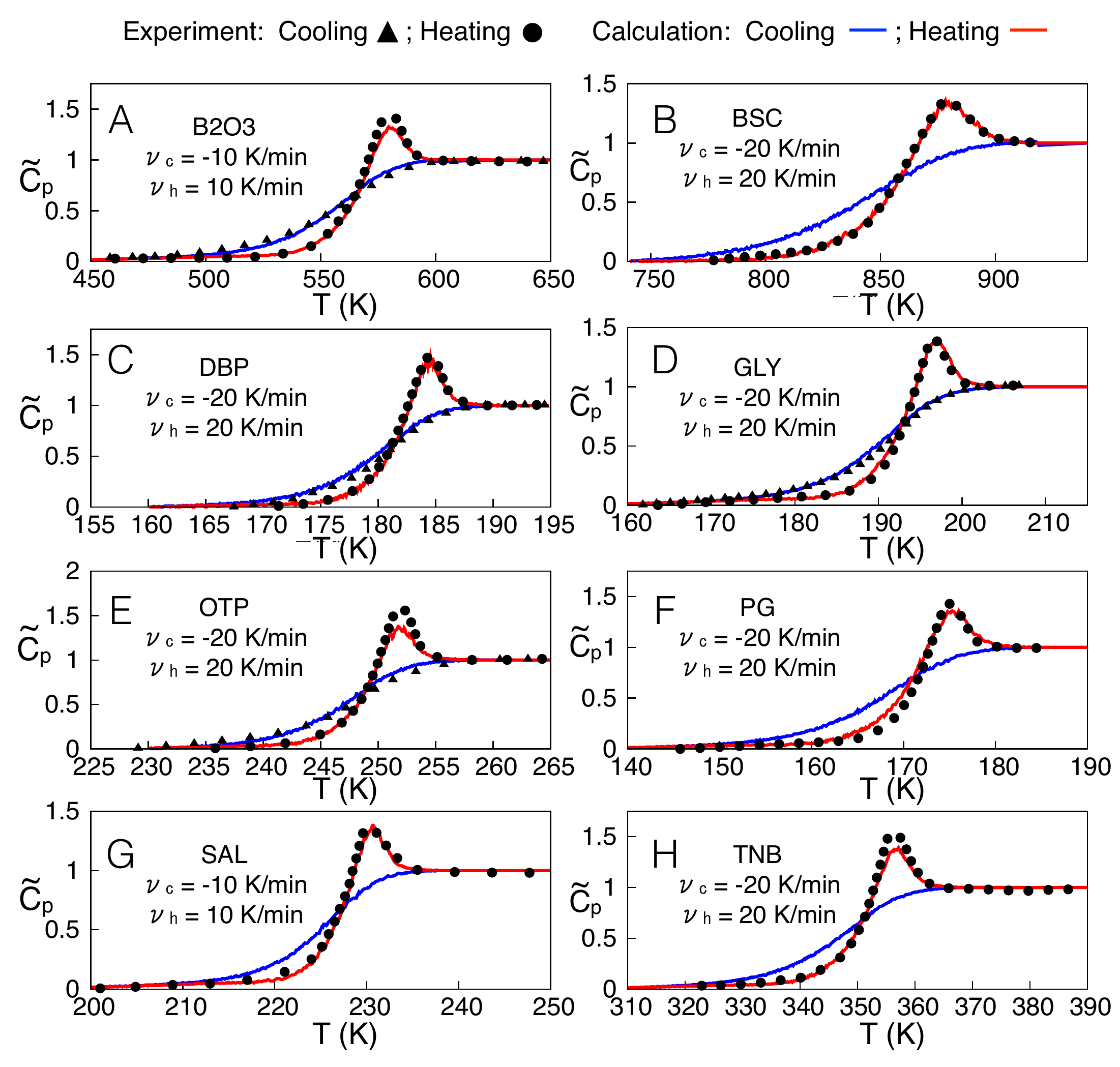}}
\caption{\label{fig:fig4}  {Comparison with heat capacity data obtained through DSC.  Panels ({\it{A}}) through ({\it{H}}) show the reduced heat capacity predicted by our model compared with experimental data.  Experimental data are depicted by points, with triangles for cooling scans (where available) and circles for heating scans.  Red and blue lines represent heating and cooling curves obtained from our calculation.  A summary of systems and parameters is given in Table~1.}}
\end{figure}

\begin{table}[t]
\fignumfont{Table~1. }\figtextfont{Material parameters for systems considered in Figs.~\ref{fig:fig4} and \ref{fig:fig5}.}
\tabletextfont
\begin{center}
\begin{tabular}{ l | r r r | c c}
 system & $T_\mathrm{o}^{a}$  & $(J/T_\mathrm{o})^{a, b}$  &  $\ln(\tau_\mathrm{o})^{a}$ & $\gamma^{c}$ & $(\sigma/\ell_\mathrm{o})^d$ \\
 \hline
boron oxide (B2O3)$^e$ & 1066 & 5.0 & -19.0 & 0.293 & 0.8 \\
\parbox[c]{2.5cm}{borosilicate crown  $\quad$  glass (BSC)$^f$}  & 2002 & 3.5 & -20.2 & 0.275 & 1.0\\
dibutylphthalate (DBP)$^g$  & 340 & 6.2 & -27.1 & 0.221 & 2.4\\ 
glycerol (GLY)$^h$ & 338 & 6.2 & -17.7 & 0.220 & 2.5\\ 
$o$-terphenyl (OTP)$^i$ & 341 & 12.9 & -20.5 & 0.244 & 1.6\\
propylene glycol (PG)$^g$ & 321 & 5.2 & -17.7 & 0.237 & 1.8 \\
salol (SAL)$^j$ & 308 & 12.6 & -19.6 & 0.275 & 1.0 \\
\parbox[c]{2.5cm}{trisnaphthylbenzene (TNB)$^k$} & 510 & 10.8 & -21.2 & 0.247 & 1.5
\end{tabular}
\end{center}
\begin{minipage}{0.49\columnwidth}
\footnotetext{$^a$Reproduced from Ref.~\cite{elmatad2009corresponding}.} 
\footnotetext{$^b$Reported in log base e.} 
\footnotetext{$^c$Best fits to calorimetry data.  See {\it Methods}.} 
\footnotetext{$^d$Reduced lengths implied by $\gamma$ with \\ $\gamma_0=0.275$. See {\it Methods}.} 
\footnotetext{$^e$Heat capacity scans obtained from Ref.~\cite{debolt1976analysis}.} 
\end{minipage}
\begin{minipage}{0.49\columnwidth}
\footnotetext{$^f$Heat capacity scans obtained from Ref.~\cite{moynihan1976dependence}.} 
\footnotetext{$^g$Heat capacity scans obtained from Ref.~\cite{wang2009enthalpy}.} 
\footnotetext{$^h$Heat capacity scans obtained from Ref.~\cite{wang2002direct}.} 
\footnotetext{$^i$Heat capacity scans obtained from Ref.~\cite{velikov2001glass}.} 
\footnotetext{$^j$Heat capacity scans obtained from Ref.~\cite{laughlin1972viscous}.} 
\footnotetext{$^k$Heat capacity scans obtained from Ref.~\cite{whitaker1996synthesis}.} 
\end{minipage}
\end{table}

\begin{figure}
\centerline{\includegraphics[width=0.4\textwidth]{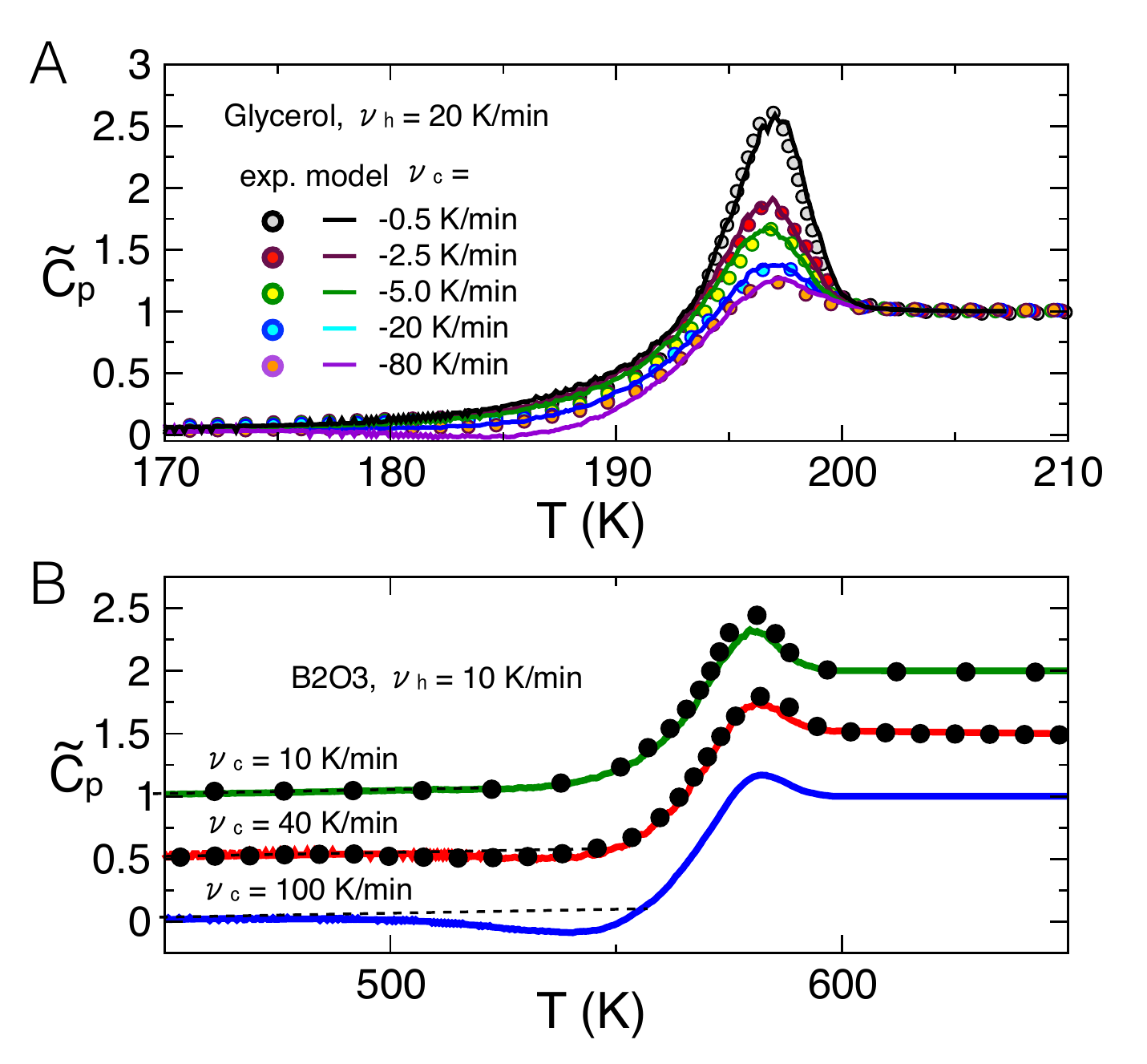}}
\caption{\label{fig:fig5}  {Behavior of the heat capacity upon heating from the glass.  Panel ({\it{A}}) depicts heating scans for a typical system (glycerol) for constant heating rate $\nu_\mathrm{h}$ and variable cooling rate, $\nu_\mathrm{c}$.  Experimental data and simulated curves are depicted by points and lines, respectively.  Slower cooling yields a more stable glass and a more prominent peak in the heat capacity on heating.  Panel ({\it{B}}) depicts heating cycles for a fixed cooling rate for a different system, $\mathrm{B}_2\mathrm{O}_3$.  When the magnitude of the heating exceeds the cooling rate, the heat capacity dips below the glassy heat capacity before recovering.  The extrapolated heat capacity of each glass is depicted by a dashed black line.  The bottommost data set with no corresponding experimental data points is a prediction of the model.  Curves are offset for clarity. }}
\end{figure}

Panels {\it{A}}-{\it{H}} of Figure~\ref{fig:fig4} compare theory and experiment for eight typical glass-formers in experiment.  Where available, standard scans~\cite{velikov2001glass} are considered, with the cooling and heating rate equal to 20~K/min.  Given $J$, $T_\mathrm{o}$ and $\tau_\mathrm{o}$ determined from equilibrium structural relaxation data~\cite{elmatad2009corresponding}, the value of $\gamma$ is obtained iteratively by optimizing fits to the calorimetry data.  The corresponding cooling and heating curves obtained from our model are depicted by blue and red lines, respectively in Fig.~\ref{fig:fig4} .  Optimizing $\gamma$ proves important for quantitative agreement between theory and experiment.  But employing a single typical value, $\gamma \approx 0.25$ for all systems considered, yields qualitative agreement between theory and experiment. Fits of comparable quality to those shown in Fig.~\ref{fig:fig4} can be obtained from TNM-like models, but requiring at least four different adjustable parameters~\cite{hodge1994enthalpy} for each system considered. 

Our East-like modeling also produces correct quantitative behavior for non-standard cooling and heating rates.  Panel~{\it{A}} of Fig.~\ref{fig:fig5} shows DSC heating scans for glycerol for glasses created at several different cooling rates.  Model curves are obtained using the parameters of Table~1 and depicted by lines.  Slower cooling yields a more stable glass with larger domains and thus a more prominent peak in the heat capacity on heating.  Panel~{\it{B}} of Fig.~\ref{fig:fig5} shows heating scans for a glass formed at a single cooling rate for $\mathrm{B}_2\mathrm{O}_3$.  When the magnitude of the heating rate exceeds the cooling rate, the system is able to further equilibrate during heating and the heat capacity dips below the heat capacity of the glass, depicted by dashed black lines, before recovering.  This behavior is particularly apparent for very large values of $| \nu_\mathrm{h}  / \nu_\mathrm{c} |$, as illustrated by the predicted data set for $\nu_\mathrm{c} = -100$~K/min.

A final qualitative behavior of interest regards the sharpness the heat capacity to drop at $T_\mathrm{g}$, which increases with decreasing cooling rates.  Here, we define $T_\mathrm{g}$ as the temperature where $\mathrm{d} \tilde{C}_\mathrm{p} / \mathrm{d} T$ is maximum upon cooling.  Empirically, the value of this non-equilibrium $[\mathrm{d} \tilde{C}_\mathrm{p} / \mathrm{d} \ln T]_{T=T_\mathrm{g}}$ correlates with the fragility of the equilibrium relaxation time, $m = -(d \ln \tau /  d \ln T) |_{T=T_\mathrm{g}}$~\cite{moynihan1993correlation}.  The East-like models exhibit this correlation too.  Specifically, over the range of typical values for $\sigma / \ell_0$, $J$, $T_\mathrm{o}$ and $\tau_\mathrm{o}$, East-like models yield $[\mathrm{d} \tilde{C}_\mathrm{p} / \mathrm{d} \ln T]_{T=T_\mathrm{g}} \approx 0.23 \,m$.  
The linear relationship for East-like models holds when relaxation is dominated by hierarchical dynamics.  It diminishes in accuracy in the regime of relatively small fragility, $m\lesssim30$, where both mean-field and hierarchical dynamics play significant roles. The proportionality constant in the linear relationship also varies with $\sigma/\ell_0$ when this ratio differs by more than an order of magnitude from 1.

\subsection{Other experiments}
Although we have focused here on configurational enthalpy and heat capacity, our model could be applied to a host of other non-equilibrium phenomena.  For example, specific volume and refractive index could be investigated with minor modifications.  More complex behaviors such as the calorimetry of remarkably stable glass~\cite{swallen2007organic} or the  uptake of water vapor~\cite{dawson2009highly, koop2011glass} might potentially be considered as well.

The prediction of large non-equilibrium correlation lengths deserves both experimental and computational investigation.  Ref.~\cite{keys2011excitations} prescribes a simulation scheme for identifying excitations (or soft spots) in the equilibrium melt based on irreversible particle displacements, which could conceivably allow for the direct measurement of a correlation length.  However, the relative scarcity of particle displacements in the glass could render such methods intractable.  An alternative method to identify excitations in the absence of their dynamics might involve localized soft modes~\cite{ashton2009relationship, chen2011measurement}.  Spatial correlations among soft regions will be reflected in the vibrational spectra of glass, which might then be used to infer a correlation length~\cite{wyart2007geometric}.  

Experimental attempts to measure the non-equilibrium lengths might involve small-angle x-ray and neutron scattering.  The distribution of these lengths reflects an anti correlation between excitations that is present in the glass but absent in the liquid.  It should be manifested in the rate at which the structure factor approaches the compressibility value as wave vector $q \rightarrow 0$~\cite{hopkins2012nonequilibrium}. Detection will require sensitivity to variations of molecular density consistent with variations in fictive-temperature field and the coefficient of thermal expansion.

\begin{materials} Our East-like model calculations are done on a $d=1$ lattice with lattice spacing $\ell_0$.  \textcolor{black}{At  temperatures relevant to experimental glass transitions, straightforward numerical simulation of the East model is impractical, but for $d=1$ we may exploit Sollich and Evans super-domain analysis~\cite{sollich2003glassy}.  A super-domain analysis might be possible in higher dimensions, but such a method is not yet known.  Rather, to compare with experiment, we map the $d=1$ results obtained in this way to a} $d=3$ system with characteristic length $\sigma$.  This {\it Methods} section provides details on the calculations and the mapping.

\section{Conversion to lattice units}
Structural relaxation occurs on a characteristic length scale, typically a molecular diameter or the length $2\pi / q_0$, where $q_0$ is the wave vector for the main peak of the structure factor.  As a result, the parameter $J$ obtained from structural relaxation data is directly related to an excitation activation energy on this length scale, $\sigma$~\cite{keys2011excitations}.  In contrast, the principal length of an East-like model is a lattice spacing, $\ell_0$.  As a result of East-like scaling, $J_\ell = J_0[1+\gamma_0 \ln(\ell/\ell_0)]$, Eq.~\eqref{eq:Jell} follows with
\begin{equation}
\label{eq:Jo}
J_0 = \frac{J_\mathrm{\sigma} }{1+ \gamma_0 \ln \left( \sigma / \ell_0 \right)}\,,
\end{equation}
and
\begin{equation}
\label{eq:gamma}
\gamma = \frac{\gamma_0}{1 + \gamma_0 \ln \left( \sigma / \ell_0 \right) }.
\end{equation}
With these equations, the energy scales of the underlying East-like model are determined in terms of $\gamma$ and $J_\sigma=J \sqrt{d_\mathrm{f} /\gamma \,\,}$.  The value of $\gamma$ accounts for entropy of relaxation pathways connecting excitations, which changes with length scale.  Here, we are concerned with the large-length regime, which applies to structural relaxation.  For the $d=1$ East model, $\gamma$ is $1/\ln 2$ at small length scales, and it is $1/2\ln2$ for lengths larger than $1/c$~\cite{chleboun2012time}. In practice, $\gamma$ (or equivalently $\sigma / \ell_0$) can be determined by fitting calorimetry data.  The equilibrium excitation concentration is given in reference lattice units by 
\begin{equation}
c_\mathrm{eq} \ell_0^d = \exp(-\tilde{\beta}\,J_0)\,, \quad \tilde{\beta} = 1/T-1/T_\mathrm{o}\,.  
\end{equation}
The length of equilibration in the cooling time scale $\Delta t$ is given by
\begin{equation}
\label{eq:DeltaEll2}
\Delta \ell / \ell_0  = \left( \Delta t / \tau_\mathrm{o} \right)^{1/\tilde{\beta} \gamma_0 J_0}.
\end{equation}

\section{Mapping between 1d and 3d}
Regardless of dimensionality, structural relaxation is dictated by the distance between excitations, $\ell$ . Inter-excitation distances (rather than excitation concentrations) are conserved when mapping between the $d=1$ and $d=3$ systems.  For $d>1$, relaxation between excitations occurs on irregular paths, so that the average length between excitations is given by $\ell_\mathrm{eq}/\ell_0 = \exp(\tilde{\beta} J_0/d_\mathrm{f})$.  The equilibrated regions of the one-dimensional proxy lattice are populated with excitations at the concentration $1/\ell_\mathrm{eq}$.

The fictive temperature profile along a line connecting excitations in an East-like model should be roughly identical for the $d=3$ and $d=1$ systems.  However, for $d=3$, the domains are space-filling, and the average fictive temperature must be re-weighted with respect to the $d=1$ case.  A mapping of the spatially-averaged fictive temperature from the $d=1$ to the $d=3$ case, that accounts for these facts is 
\begin{equation}
\label{eq:Tf3}
T_\mathrm{f} = \sum\limits_{D \in \{ D \} } [N(D)]^2 \sum\limits_{i \in D } T_{\mathrm{f}, i} \ \ / \sum\limits_{ D \in \{D\}  }[N(D)]^3 .
\end{equation} 
Here, $D$ is set of proxy lattice sites $i \cup \{j \in N : i < j \leq i+N(D)-1\}$ satisfying $n_i=1$, $n_j=0$ and $n_{i+N(D)} = 1$.  The set of all such domains within the system is denoted $\{D\}$ and $N(D)$ represents the number of lattice sites in domain $D$.  We find that $T_\mathrm{f}$ is reasonably insensitive to details of the method chosen to mimic three-dimensional domains with one-dimensional domains treated in the calculation.  For example, constructions where excitation fronts are assumed to move forward in a cone prove equally satisfactory \textcolor{black}{(see SI)}.  The relative insensitivity to this detail is a result of the directional facilitation in East-like models.

\section{Fictive temperature computation algorithm}
The persistence field for a given system is propagated by running dynamics of the proxy $d=1$ East-like model at a temperature corresponding to equilibrium concentration $1/\ell_\mathrm{eq}$.  This is done for a time $\Delta t$ corresponding to the time scale for the equilibration length $\Delta \ell$ computed from Eq.~\eqref{eq:DeltaEll2}.  At each step $k$, the persistence field is computed, the fictive temperature field is updated, and the average fictive temperature is computed from Equation~\eqref{eq:Tf3}.  The process is then repeated with the new working temperature for step $k+1$.  The computation is decomposed into small enough temperature steps to negate quadrature error.  We find that using 1000 steps is sufficiently large to accurately approximate time dependence for the non-equilibrium protocols considered here.

Standard Monte-Carlo simulation of East-like models becomes prohibitively expensive when $\tilde{\beta}J_0 > 3$.  The non-equilibrium systems we consider to compare with experiment are typically at $\tilde{\beta}J_0 > 5$.  The ``superdomain'' construction of Sollich and Evans~\cite{sollich1999glassy, sollich2003glassy} for computing persistence fields of East-like models is accurate and efficient in this region, and we therefore employ this procedure.  The superdomain construction derives from that fact that at low enough temperatures, relaxation in East-like models can be partitioned in terms of relaxation on different scales, and each separate scale is nearly independent of the others. Although the original Sollich-Evans method is designed to operate at a fixed temperature, we make a small modification to allow for multiple temperature steps.  

At the end of each step $k$, the system is left with a set of non-equilibrated excitations called ``superpspins.''  Each superspin possesses an  equilibrated zone of length $\Delta \ell$, adjacent to the spin in the direction of facilitation.  Excitations within these equilibrated zones are inconsequential in the Sollich-Evans procedure and therefore not kept track of.  An initial configuration for $k+1$ cooling or heating step is generated from the superspins of step $k$ by populating excitations within equilibrated zones.  The positions of these excitations are drawn at random from the Poisson distribution with equilibrium concentration $1/\ell_\mathrm{eq}$. 
\textcolor{black}{Each calculation reported here is averaged over 10,000 trajectories for a system of size N=32,~768.  The statistical error is less than or equal to the width of the lines used to represent the data.  A user-friendly program for carrying out these calculations is found at \url{}{http://www.nottingham.ac.uk/\%7eppzjpg/GLASS}.}

\end{materials}

\begin{acknowledgments}
We thank D.T. Limmer, S. Vaikuntanathan, and Y.J. Jung for insightful discussions.  This work was supported by the Director, Office of Science, Office of Basic Energy Sciences, and by the Division of Chemical Sciences, Geosciences, and Biosciences of the U.S. Department of Energy at LBNL, and by the Laboratory Directed Research and Development Program at Lawrence Berkeley National Laboratory under Contract No. DE-AC02-05CH11231.  
\end{acknowledgments}

\bibliographystyle{pnas}

\end{article}

\end{document}